\documentstyle[amssymb,eqsecnum,aps,multicol,epsf]{revtex}

\begin{document}
\draft
\title{Spectrum of light scattering from an extended atomic wave packet}
\author{B. Dubetsky and P. R. Berman}
\address{Michigan Center for Theoretical Physics}
\address{Physics Department, University of Michigan, Ann Arbor, MI 48109-1120}
\date{\today}
\maketitle

\begin{abstract}
The spectrum of the light scattered from an extended atomic wave packet is
calculated. For a wave packet consisting of two spatially separated peaks
moving on parallel trajectories, the spectrum contains Ramsey-like fringes
that are sensitive to the phase difference between the two components of the
wave packet. Using this technique, one can establish the mutual coherence of
the two components of the wave packet without recombining them.
\end{abstract}

\pacs{32.80.-t, 39.25 +k, 52.38.Bv}

\preprint{HEP/123-qed}

\section{Introduction}

The typical operation of a matter-wave interferometer \cite{c1,berm}
involves a beam splitter that separates an incoming atomic beam into a set
of states having different center-of-mass momenta, and mirrors which
recombine the beams. The atom density of the recombined beams exhibits
interference fringes resulting from different phases acquired during free
motion along the different arms of the interferometer. In this paper, we
address a question of fundamental importance, {\it Is it possible to
establish the spatial coherence of the wave packet without recombining the
beams?}

To accomplish this goal, we propose a method involving Rayleigh scattering.
It has already been shown \cite{c6} that the scattered signal, {\em %
integrated over frequency}, is not sensitive to the spatial coherence of
atom wave packets. A similar conclusion was reached for the correlation
properties of the field emitted in spontaneous emission \cite{paul}. On the
other hand, it was shown that the {\em spectrum }of spontaneous emission 
{\em was} sensitive to the spatial form of the wave packet \cite{c8}. In
effect, if one measures a frequency integrated spectrum, information on the
momentum distribution of the atom wave packet is lost. The scattering cross
section is then a sum over contributions \ from each position in the wave
packet, with no interference. On the other hand, a measure of the scattered
spectrum is equivalent to a measure of the momentum of the atom and all
information on the position is lost. In this way contributions from
different spatial positions can interfere. Similar conclusions can be
reached on the modification of fringe contrast in an atom interferometer
resulting from light scattering, but the localization in that case is on the
order of a wavelength \cite{c6,pritch,mlynek}.

It is shown below that the spectrum of radiation scattered from the atom
wave packets in two arms of an interferometer allows one to probe the
spatial coherence between the beams. In other words, it is not necessary to
recombine the beams to observe the interference between the beams. For a
two-peaked initial wave packet, the scattered signal, as a function of the
frequency difference between incident and scattered fields, exhibits a type
of Ramsey fringe structure.

\section{Spectrum}

We consider the scattering of classical radiation by an atom having
center--of-mass wave function $\psi \left( {\bf r}\right) $ and internal
state wave function $\psi _{g}.$ Radiation from a mode $\left\{ {\bf k}%
_{i},\omega _{i}\right\} $ of the incident field is scattered into a mode $%
\left\{ {\bf k}_{f},\omega _{f}\right\} $ of the vacuum field$.$ Scattering
occurs via an off-resonant intermediate internal state $n$ of the atom. The
atom remains in its initial internal state following the scattering, but the
momentum of the atom changes from ${\bf p}$ to $\left( {\bf p-}\hbar {\bf q}%
\right) ,$ where ${\bf q=k}_{f}-{\bf k}_{i}$. If the momentum state wave
function is denoted by $\Phi ({\bf p}),$ then the scattering cross section
is given by the Kramers-Heisenberg expression 
\begin{mathletters}
\label{1}
\begin{equation}
d\sigma =\frac{d\sigma }{d{\bf n}_{f}}P\left( \Delta \right) d\omega _{f}d%
{\bf n}_{f},  \eqnum{2.1}  \label{1b}
\end{equation}
where 
\end{mathletters}
\begin{equation}
P\left( \Delta \right) =\int d{\bf p}\left| \Phi ({\bf p})\right| ^{2}\delta
\left( \Delta +\omega _{{\bf q}}-{\bf q\cdot p/m}\right) ,  \label{2}
\end{equation}
and 
\begin{equation}
\frac{d\sigma }{d{\bf n}_{f}}=\frac{\omega _{i}\omega _{f}^{3}}{\hbar
^{2}c^{4}}\left| \sum_{n}\left\{ \frac{\left( {\bf d}_{gn}\cdot {\bf e}%
_{f}^{\ast }\right) \left( {\bf d}_{ng}\cdot {\bf e}_{i}\right) }{\omega
_{ng}-\omega _{i}}+\frac{\left( {\bf d}_{gn}\cdot {\bf e}_{i}\right) \left( 
{\bf d}_{ng}\cdot {\bf e}_{f}^{\ast }\right) }{\omega _{ng}+\omega _{f}}%
\right\} \right| ^{2}  \label{21}
\end{equation}
In these expressions, $\Delta =\omega _{f}-\omega _{i}$ is the frequency
detuning between scattered and incident field modes, ${\bf n}_{f}={\bf k}%
_{f}/k_{f},$ ${\bf d}_{gn}$ is a dipole moment matrix element, ${\bf e}_{f}$
and ${\bf e}_{i}$ are polarization vectors, and $\omega _{{\bf q}}=\hbar 
{\bf q}^{2}/2m$ is a recoil frequency associated with the change in atomic
momentum $\hbar {\bf q.}$ The center-of mass energy has been neglected in
the denominators in Eq. (\ref{21}). The function $P\left( \Delta \right) $
determines the scattering spectrum. Since $\int P\left( \Delta \right)
d\Delta =1$, the {\em integrated} spectrum does not depend on the form of
the center-of-mass wave function, in agreement with Ref. \cite{c6}.

For a one-dimensional wave packet consisting of a superposition of two
identical packets $\psi _{a}\left( x\right) $ having extent $a$, phase
difference $\phi ,$ distanced from one another by a large distance $L\gg a,$
and having relative momentum $p_{0}$, i. e. $\psi \left( x\right) =2^{-1/2}%
\left[ \psi _{a}(x)+e^{-i\phi }e^{ip_{0}x/\hbar }\psi _{a}(x-L)\right] ,$
the wave function in momentum space is given by 
\begin{equation}
\Phi (p)=2^{-1/2}\left\{ \Phi _{a}(p)+\Phi _{a}(p-p_{0})\exp \left[ -i\left(
\phi +\left( p-p_{0}\right) L/\hbar \right) \right] \right\} ,  \label{41}
\end{equation}
where $\Phi _{a}(p)$ is the Fourier transform of $\psi _{a}\left( x\right) $
and it has been assumed that $\psi _{a}\left( x\right) $ is real and an even
function of $x$. The spectrum is given by 
\begin{eqnarray}
P\left( \Delta \right) &=&\frac{m}{2q_{x}}\left\{ \left| \Phi _{a}\left[
\left( \Delta +\omega _{q}\right) m/q_{x}\right] \right| ^{2}+\left| \Phi
_{a}\left[ \left( \Delta +\omega _{q}\right) m/q_{x}-p_{0}\right] \right|
^{2}\right.  \nonumber \\
&&\left. +2\Phi _{a}\left[ \left( \Delta +\omega _{q}\right) m/q_{x}\right]
\Phi _{a}\left[ \left( \Delta +\omega _{q}\right) m/q_{x}-p_{0}\right] \cos %
\left[ \phi +\left( \Delta +\omega _{q}\right) Lm/\hbar q_{x}-p_{0}L/\hbar %
\right] \right\}  \label{5}
\end{eqnarray}

In the three-dimensional case, one can choose a double-peaked, Gaussian
packet 
\begin{mathletters}
\label{e3}
\begin{equation}
\psi \left( {\bf r}\right) =2^{-1/2}\left( 2/\pi a^{2}\right) ^{3/4}A\left\{
e^{-{\bf r}^{2}/a^{2}}+e^{i{\bf p}_{0}\cdot {\bf r}/\hbar -i\phi }e^{-\left( 
{\bf r-L}\right) ^{2}/a^{2}}\right\} ,  \eqnum{2.6}  \label{6}
\end{equation}
where 
\end{mathletters}
\begin{equation}
A=\left[ 1+\cos \left( \phi -\frac{{\bf p}_{0}\cdot {\bf L}}{2\hbar }\right)
\exp \left( -\frac{L^{2}}{2a^{2}}-\frac{p_{0}^{2}a^{2}}{2\hbar ^{2}}\right) %
\right] ^{-1/2}  \label{61}
\end{equation}
is a normalization factor [Eqs. (\ref{6}) and (\ref{61}) are valid for
arbitrary ratios of $L/a$].. For this packet one finds 
\begin{eqnarray}
P\left( \Delta \right) &=&\frac{amA^{2}}{2\sqrt{2\pi }\hbar q}\left\{ \exp %
\left[ -\left( \frac{m\left( \Delta +\omega _{q}\right) }{q}\right) ^{2}%
\frac{a^{2}}{2\hbar ^{2}}\right] +\exp \left[ -\left( \frac{m\left( \Delta
+\omega _{q}\right) }{q}-\frac{{\bf p}_{0}\cdot {\bf q}}{q}\right) ^{2}\frac{%
a^{2}}{2\hbar ^{2}}\right] \right.  \nonumber \\
&&+2\exp \left[ -L^{2}\sin ^{2}\left( \theta \right) /2a^{2}-\left( \frac{%
m\left( \Delta +\omega _{q}\right) }{q}\right) ^{2}\frac{a^{2}}{2\hbar ^{2}}%
-\left( p_{0}^{2}+\left( {\bf p}_{0}\cdot {\bf \hat{q}}\right) ^{2}\right) 
\frac{a^{2}}{8\hbar ^{2}}-\frac{m\left( \Delta +\omega _{q}\right) }{q}%
\left( {\bf p}_{0}\cdot {\bf \hat{q}}\right) \frac{a^{2}}{2\hbar ^{2}}\right]
\nonumber \\
&&\times \left. \cos \left[ \phi -\frac{{\bf p}_{0}\cdot {\bf L+}\left( {\bf %
p}_{0}\cdot {\bf \hat{q}}\right) \left( {\bf L\cdot \hat{q}}\right) }{2\hbar 
}+\frac{\left( \Delta +\omega _{q}\right) mL\cos \left( \theta \right) }{%
\hbar q}\right] \right\}  \label{7}
\end{eqnarray}
where ${\bf \hat{q}=q/}q$ is a unit vector along ${\bf q}$ and $\theta $ is
the angle between ${\bf q}$ and ${\bf L.}$

\section{Discussion}

It is clear from Eq. (\ref{2}) that the spectrum is simply the momentum
distribution of the entire wave packet, evaluated at momenta determined by
the resonance condition 
\begin{equation}
{\bf q\cdot p/}m{\bf =}\Delta {\bf +}\omega _{{\bf q}}.  \label{8}
\end{equation}
For a double peaked wave function, the momentum distribution oscillates as a
function of ${\bf p}$, and this oscillation can be mapped into the spectrum
of the scattered radiation. For the interference term to contribute, it is
necessary that $\frac{p_{0}a}{\hbar }\lesssim 1$. Let us see how this
condition applies in an atom interferometer.

A well-collimated atomic beam is incident on a beam splitter that splits the
beam into two momentum components. We can imagine that a momentum difference
in the $x$ direction, $p_{0}=2\hbar k,$ is produced via frequency controlled
Bragg scattering from two counterpropagating fields \cite{bian} or some
equivalent process. The quantity $k=2\pi /\lambda $ is the field propagation
constant. For the Bragg field to split the beam into two distinct packets it
is necessary that $p_{0}t/m=L\gg a$, and for negligible spreading of the
packet, it is necessary that $\left( \hbar /ma\right) t\ll a$. These
conditions can be satisfied simultaneously only if $p_{0}a/\hbar \gg 1.$ In
other words, if the Bragg field splits the incident beam into two, separated
beams, the scattering techniques described above cannot be used to reveal
the coherence of the wave packet since the interference term vanishes! A way
around this is to apply a second Bragg pulse after the beams are separated.
By a proper choice of the Bragg pulse it is possible to return the relative
velocity of the split beams to a value $p_{0}=0.$ The beams will still be
spatially separated, but moving on parallel trajectories. As such, one can
analyze scattering from a two-peaked packet when the relative momentum ${\bf %
p}_{0}=0.$

With ${\bf p}_{0}=0$, Eq. (\ref{7}) reduces to 
\begin{equation}
P\left( \Delta \right) =\frac{amA^{2}}{\sqrt{2\pi }\hbar q}\exp \left[
-\left( \frac{m\left( \Delta +\omega _{q}\right) }{q}\right) ^{2}\frac{a^{2}%
}{2\hbar ^{2}}\right] \left\{ 1+\exp \left[ -L^{2}\sin ^{2}\left( \theta
\right) /2a^{2}\right] \cos \left[ \phi +\frac{\left( \Delta +\omega
_{q}\right) mL\cos \left( \theta \right) }{\hbar q}\right] \right\}
\label{91}
\end{equation}
Owing to recoil effect \cite{c7} the spectrum as a whole is shifted from $%
\Delta =0$ by the recoil frequency $-\omega _{q}$. The spectral width $%
\gamma $ of the envelope of the signal is of order 
\begin{equation}
\gamma \sim \omega _{q}/qa  \label{10}
\end{equation}
The interference of the two momentum state wave packets represented in Eq. (%
\ref{91}) translates in frequency space into oscillations having period 
\begin{equation}
\gamma _{R}=4\pi \omega _{q}/qL\cos \left( \theta \right) .  \label{11}
\end{equation}
To observe these oscillations, one must have $\gamma _{R}<\gamma $ or,
equivalently, $L\cos \left( \theta \right) >a$. The oscillations in
frequency space have the same structure encountered in Ramsey fringes. The
central fringe occurs for $\Delta =-\omega _{q}-\left[ \hbar q/Lm\cos \left(
\theta \right) \right] \phi $. If, instead of a coherent superposition of
two spatially separated wave packet components, one had chosen an {\em %
incoherent} sum of two separated wave packets, the interference term would
be absent (corresponding to an average over $\phi $). Thus, in principle,
one can establish the mutual spatial coherence of the spatially separated
wave packet components without recombining them.

A similar effect has been predicted previously in the spectrum of
spontaneous emission from an extended wave packet. \cite{c8}. However, in
order to resolve the effects related to the size of the wave packet in that
case, it is necessary that the width of the atomic wave function in momentum
space $\delta p\sim \hbar /a$ be larger than $m\Gamma \lambda $, where $%
\Gamma $ is the upper state decay rate and $\lambda $ is the wavelength of
the transition. This requirement restricts the wave packet size to $%
a\lesssim \left( \omega _{k}/\Gamma \right) \lambda ,$ which is typically
much smaller (10$^{-2}$ to 10$^{-3}$) than an optical wavelength.

In our case, the size of the wave packet is limited only by the requirement
that the wave packet be coherent over a distance $a.$ The spectral
resolution needed to observe the coherence effects represented in Eq. (\ref
{91}) is of order $\omega _{q}/qa\sim \left( \lambda /4\pi a\right) \omega
_{2k}$ for backward scattering when $q=2k.$ Using a well-collimated atomic
beam or a released Bose condensate \cite{c5}, one finds that a resolution of
order $2\pi \times \,1.0\,$kHz is needed. The experimental challenge is
great to say the least. The scattered signal can be detected using
heterodyne techniques, but the signal strength is small, the collection
angle is small, and long integration times can be anticipated. The signal to
noise can be improved if, instead of measuring the scattered spectrum, one
adds a probe field and monitors the probe field absorption or index change
as a function of probe-pump field detuning.

It may also be possible to reduce the resolution requirements by considering
scattering from a {\em multicomponent} wave packet rather than a two-peaked
wave packet. For example, if one scatters an atomic beam from a resonant
standing wave field (resonant Kapitza-Dirac effect\cite{c9}) or from a
microfabricated structure having grating period ${\bf d}$, the momentum
space density, $\left| \Phi ({\bf p})\right| ^{2},$ following the
interaction consists of a set of narrow peaks centered at ${\bf p=}n\hbar 
{\bf q}_{s}{\bf ,}$ where ${\bf q}_{s}{\bf =}2\pi {\bf \hat{d}/}d$ and $n$
is an integer whose maximum value is determined by the strength of the
interaction. For this momentum distribution, one finds from Eq. (\ref{2})
that the spectrum consists of a set of peaks (recoil components) centered at 
\begin{equation}
\Delta =-\omega _{q}+n{\bf q\cdot q}_{s}{\bf /}m.  \label{12}
\end{equation}
If atoms are scattered by a standing wave and ${\bf q=q}_{s}$, the recoil
components are distanced from one another by $2\omega _{q},$ which for the $%
D_{2}$ line in Na, is equal to $2\pi \,208\,KHz.$ For $n>1$ the spectral
width is correspondingly larger. Note that it is still necessary to have a
spectral width of order $\omega _{q}/qa\sim \left( \lambda /4\pi a\right)
\omega _{2k}$ to resolve the interference pattern of Eq. (\ref{91}); the
larger resolution quoted above refers to the spectral width of the {\em %
entire} scattered signal.

\acknowledgments

We are grateful to T. Sleator and K. Rz\c{a}\.{z}ewski for encouraging
discussions. This work is supported by the U. S. Army Research Office under
Grant Nos. DAAG55-97-0113 and DAAD19-00-1-0412, and by the National Science
Foundation under Grant No. PHY-9800981.

\end{document}